\documentclass[aps,prl,twocolumn,superscriptaddress]{revtex4-1}
\usepackage{graphicx}

\begin{document}


\title{Undiscovered pulsar in the Local Bubble as an explanation of the local high energy cosmic ray all-electron spectrum}


\author{R. L\'opez-Coto}
\thanks{rlopez@pd.infn.it}
\affiliation{INFN and Universit\`{a} di Padova, I-35131, Padova, Italy}
\affiliation{Max-Planck-Institut f\"ur Kernphysik, P.O. Box 103980, D 69029 Heidelberg, Germany}

\author{R.D. Parsons}
\author{J.A. Hinton}
\author{G. Giacinti}
\affiliation{Max-Planck-Institut f\"ur Kernphysik, P.O. Box 103980, D 69029 Heidelberg, Germany}


\date{\today}

\begin{abstract}

Cosmic ray electrons and positrons are tracers of particle propagation in the interstellar medium (ISM). A recent measurement performed using H.E.S.S. extends the all-electron (electron+positron) spectrum up to 20~TeV, probing very local sources and transport due to the $\sim$10~kyr cooling time of these particles.
An additional key local measurement was the recent estimation of the ISM diffusion coefficient around Geminga performed using HAWC. The inferred diffusion coefficient is much lower than typically assumed values. It has been argued that if this diffusion coefficient is representative of the local ISM, pulsars would not be able to account for the all-electron spectrum measured at the Earth.
Here we show that a low diffusion coefficient in the local ISM is compatible with a pulsar wind nebula origin of the highest energy electrons, if a so far undiscovered pulsar with spin-down power $\sim 10^{33-34}$ erg/s exists within 30 to 80~pc of the Earth. The existence of such a pulsar is broadly consistent with the known population and may be detected in near future survey observations.
\end{abstract}

\keywords{gamma rays -- cosmic rays -- pulsars}

\maketitle

\section{Introduction}
\label{sec:intro}
The measured cosmic ray spectrum at the Earth is dominated by protons and nuclei, 
with a much smaller fraction of electrons and positrons. The electron spectrum is dominated by primary cosmic rays accelerated inside sources as supernova remnants or pulsar wind nebulae. The positron spectrum, at least below energies of $\sim$ 10~GeV, is dominated by secondary production in cosmic-ray collisions in the interstellar medium (ISM). Electrons and positrons suffer severe energy losses due to interstellar magnetic and radiation fields~\cite{Atoyan95}. That the local all-electron spectrum (LAES hereafter) provides information about very local sources and might contain a large positron fraction at high energies, has long been recognized~\cite{Atoyan95, Aharonian95}. Nevertheless, the detection of an increasing positron fraction at the highest energies 
\cite{Adriani09, Ackermann12, Aguilar14} has led to intense speculation on their origin. This positron excess (over expected secondary production) has been postulated to arise from pulsar wind nebulae \citep{Aharonian95,Hooper08}, microquasar jets \citep{Gupta14} or dark matter annihilation~\citep{Bergstrom08}. According to the most commonly accepted propagation theories within the ISM, the highest-energy positrons measured by satellites ($\sim$500~GeV) must originate from a region within $\sim$1 kpc from the Earth, limiting the number of possible sources producing this excess \citep{Aharonian95}.

Recent measurements of the LAES 
demonstrate the existence of very local sources of high energy electrons.
H.E.S.S. \cite{HESS_electrons} presented a preliminary measurement of the LAES extending up to $\sim$20~TeV with a break at $\sim$900~GeV, later confirmed by DAMPE \cite{Ambrosi17} and CALET \cite{Adriani17}. 
Furthermore, the HAWC collaboration has recently reported the diffusion coefficient inferred from the VHE $\gamma$-ray measurements of the Geminga region \cite{Geminga_hawc}. The energy-dependent diffusion coefficient is usually characterized as
$D(E)=D_0 (E/10 \mathrm{GeV})^\delta$. The HAWC inferred value is $D$(100~TeV) =  $4.5\times10^{27}$ cm$^2$/s, which assuming diffusion in Kolmogorov turbulence $\delta=0.33$ this implies $D_0 \sim 2 \times10^{26}$ cm$^2$/s. This is two orders of magnitude lower than the average of the ISM derived from Galactic propagation codes fitting primary and secondary species spectra, see e.g.~\cite{Amato_diffusion_review}.
They argue that the diffusion coefficient measured does not need to be limited to the region surrounding Geminga, but may be characteristic of the local ISM. It has been argued that if this is indeed the case, the LAES could not be explained by known nearby pulsars~\cite{Hooper_HESS17}.

\begin{figure}
\begin{center}
\includegraphics[width=0.99\columnwidth]{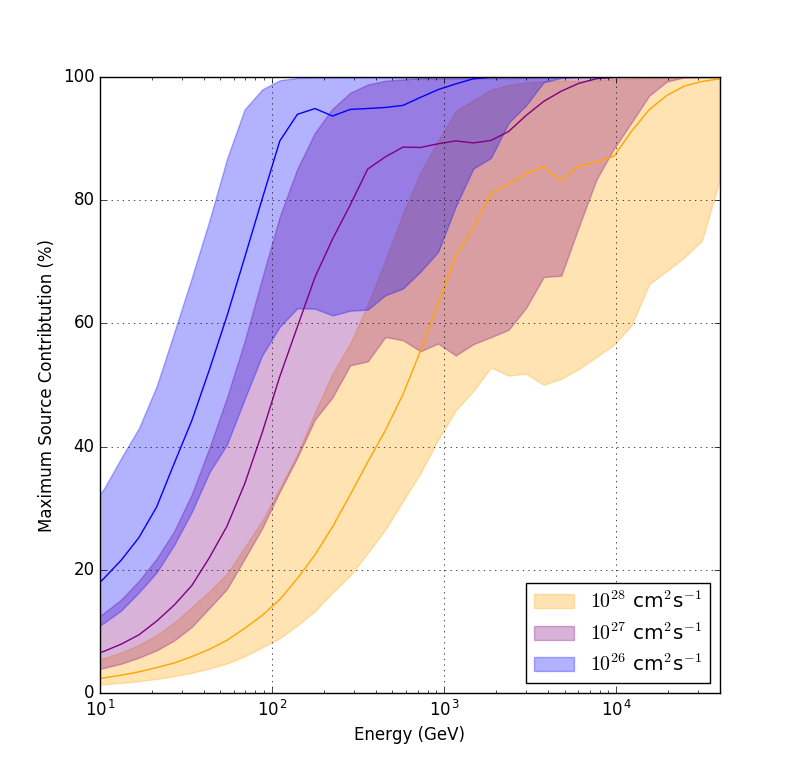}
\caption{Maximum contribution of a single source to the total cosmic ray electron spectrum as a function of energy for three diffusion coefficients with an energy dependence of $\delta=0.33$. Generated from 1000 realizations of a source population and simple propagation model. Solid lines correspond to the median of the distribution and the bands correspond to the 16$^{\rm th}$ and 84$^{\rm th}$ percentiles.}
\label{fig:npulsars_contributors}
\end{center}
\end{figure}

Whilst the GeV electron spectrum in the ISM must include contributions from a whole population of sources across the Galaxy, at energies of many TeV it becomes increasingly likely that a single source dominates. A low diffusion coefficient exacerbates this problem, confining electrons to a smaller region around their sources within their cooling time.  
Figure \ref{fig:npulsars_contributors} explores this effect quantitatively using a simple population model from the \emph{gammapy} package \cite{gammapy} assuming a supernova rate of 3 per century. There sources were distributed using the radial distribution described in \cite{Lorimer06} with the velocity distribution and spiral arms used by \cite{Faucher06}. The maximum fraction of the locally measured all electron flux contributed by a single source is shown for 
three different diffusion coefficients over 
1000 realizations of the Galaxy. 
Sources are simulated with a pulsar-like evolution (described in detail later)
and simplified particle propagation neglecting the Klein-Nishina effect in the inverse Compton losses \citep{Atoyan95}. As expected the highest energies are dominated by a small number of sources due to cooling effects. The energy at which a single source dominates shifts to lower values for lower diffusion coefficients. With the typically used diffusion coefficient of $D_0=10^{28}$\,cm$^{2}$s$^{-1}$ 
a single source dominates the spectrum in almost all realizations above 2\,TeV. With a diffusion coefficient of  $D_0=10^{26}$\,cm$^{2}$s$^{-1}$ a single source may dominate as early as 100\,GeV. 


Here we argue that even in slow diffusion scenarios a single pulsar wind nebula (PWN) explanation of the high energy LAES remains plausible. We estimate the required parameters for the pulsar powering such a PWN, its consistency with the known population and the likelihood that it could remain undetected. 



\section{Approach} 


\label{sec:theor}
We use \texttt{edge}, a code that calculates the isotropic diffusion of electrons from a point source \cite{Diffusion_electrons_paper} to compute the LAES produced by the considered source. The code computes the solution of the diffusion-loss equation using the quasi-analytical result put forward in \cite{Atoyan95}. The time-dependence of the electron injection is given by the assumption that the power source is a pulsar and its luminosity evolution is given by: $\dot{E}(t) = \dot{E}_0\left(   1 +  \frac{t}{\tau_0}\right)^{-\frac{n+1}{n-1}}$ 
where $n$ is the braking index of the pulsar and $\dot{E}_0$ the initial luminosity. $\tau_0$ is the initial spin-down timescale of the pulsar and is related to the initial birth period of the pulsar $P_0$, the current period ($P$) and the characteristic age ($\tau_{\mathrm{c}}$) by: $(P/P_0)^{n-1} = 2 \tau_{\mathrm{c}}/\tau_0(n-1)$. For this work we assume a $\tau_{\mathrm{c}}$=342 kyr and a $P=237$ ms as in the Geminga pulsar and hence $\tau_0$ is a function of the assumed $P_{0}$. We use $P_{0}$=40 ms, corresponding to $\tau_0=10^4$ yr. The spin-down power simulated is $\dot{E}=3.2\times10^{34}$ erg/s.

We consider that electrons and positrons are injected 
into the ISM from a pulsar wind nebula (PWN) with a power-law spectrum $dN/dE = f_0 E^{-\alpha}$ as appears to be the case for Geminga~\cite{Geminga_hawc} and as may be the case in general for sufficiently old systems. 
For the particle propagation we only consider diffusion with the normalization measured by HAWC $D_0=2\times10^{26}$ cm$^2$s$^{-1}$ and the energy-dependence $\delta$=0.33 corresponding to an ISM dominated by Kolmogorov turbulence. Assuming a larger $\delta$ would imply faster propagation and the best fit results would be shifted, but the dominant effect will be that of the spectral index of the electrons $\alpha$. We assume a typical magnetic field in the ISM of $B=3 \mu$G and photon energy densities for Cosmic Microwave Background (CMB), infrared (IR) and optical radiation fields of $\epsilon_{\rm{CMB}}$=0.26 eV/cm$^{-3}$, $\epsilon_{\rm{IR}}$=0.3 eV/cm$^{-3}$, $\epsilon_{\rm{Opt}}$=0.3 eV/cm$^{-3}$ for the photon targets, which are the typical values derived by GALPROP for the location of Geminga \citep{Moskalenko_Strong_electrons}. The magnetic field in the nearby ISM should be in the range between 3-5 $\mu$G according to \cite{correlation_length}. We take into account losses due to ionization, bremsstrahlung, synchrotron and inverse Compton in the Klein-Nishina regime.

\begin{figure}
\vspace{-0.5cm}
\begin{center}
\includegraphics[width=0.49\textwidth]{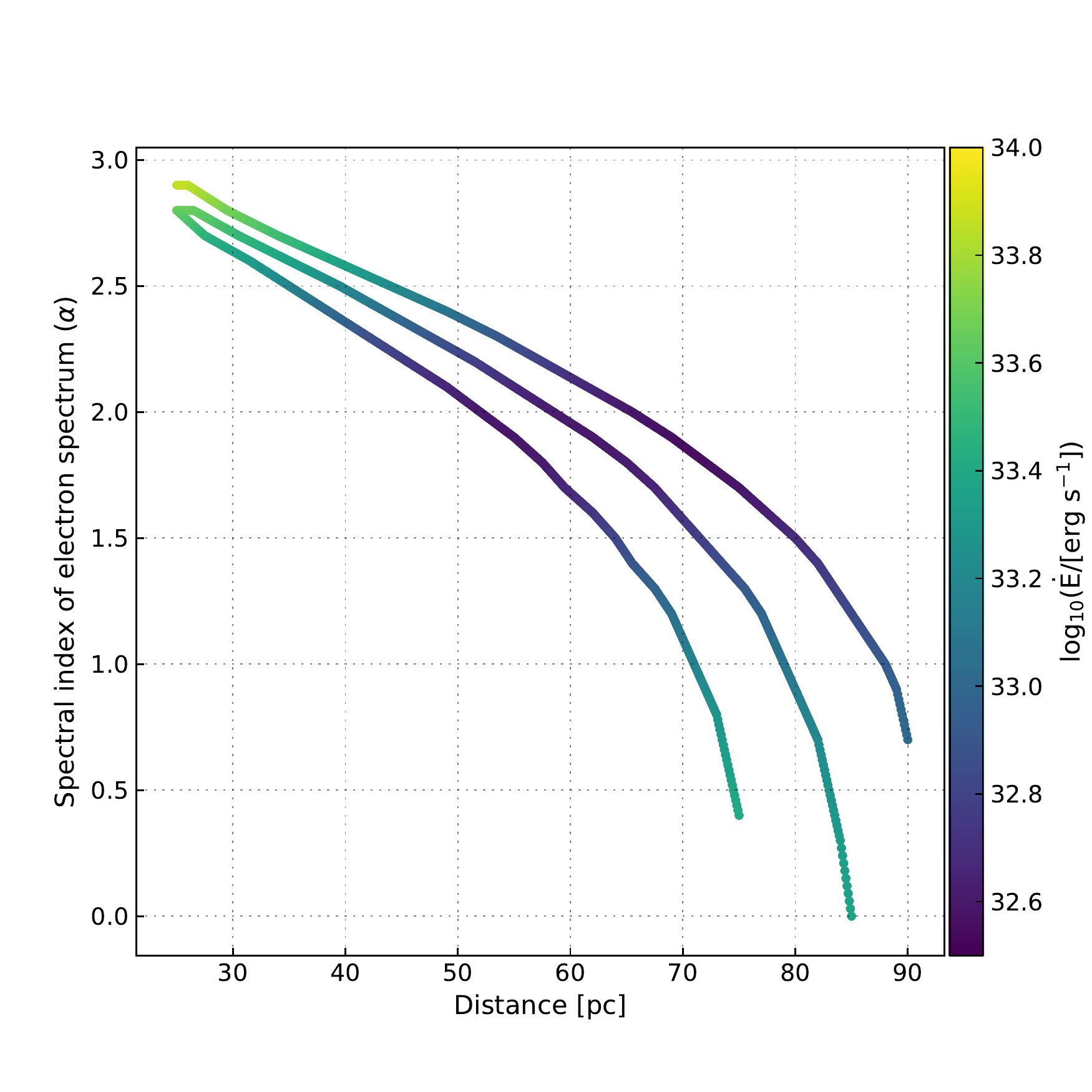}
\vspace{-1cm}
\caption{Required spectral index $\alpha$ vs pulsar distance for different magnetic fields (top line: $B=3 \mu$G, middle line: $B=4 \mu$G, bottom line: $B=5 \mu$G). The color scale gives the energy into accelerated electrons needed to match the local electron flux in the range 1--20~TeV as measured by H.E.S.S..}
\label{fig:alpha_vs_distance}
\end{center}
\end{figure}

\section{Characteristics of a single source}

Under the assumptions and calculation method given above we explore the available pulsar parameters in an attempt to match the observed LAES at TeV energies for a single source and a slow diffusion coefficient. The following are the free parameters of the fit with their allowed ranges:

\begin{itemize}
\item Distance (25~pc$<d<$100~pc)
\item Spectral index of injected electrons (0$<\alpha<$3)
\end{itemize}

The characteristic age $\tau_c$ is not a critical parameter if the equilibrium where all electrons in the 1--20~TeV regime are already cooled has been reached. Therefore, for sources with age $\gtrsim$300~kyr~ this parameter does not affect the measurements(see Figure 11 of~\cite{Diffusion_electrons_paper}). For younger pulsars, they could not account for all the LAES above 1 TeV as it is required for this diffusion coefficient by Figure \ref{fig:npulsars_contributors}. Hence for this study we fix $\tau_c$ to the 342~kyr value of Geminga~\citep{ATNF}. The amount of energy into electrons is derived by normalization to the measured flux (see below).

After calculating the LAES for a grid of the parameters in the ranges above, we fit the spectrum between 1 and 20~TeV using a single power-law and choose for each distance the $\alpha$ that produces the closest spectral index to that measured by H.E.S.S. in the same energy range. 
We consider three different magnetic field strengths ($B=3, 4$ and 5 $\mu$G) to investigate the dependence 
on synchrotron losses. The results for the best fit values for every pair of $d-\alpha$ are shown in Figure \ref{fig:alpha_vs_distance}. For every pair of values, the best fit requires a different energy into electrons $\mu\dot{E}$ to fit H.E.S.S. data in this energy range. Here $\mu$ is the fraction of the total spin-down power that is transformed into electrons. The relationship between distance and predicted slope, and hence the requirement for a different injection index, is illustrated in~\cite{Diffusion_electrons_paper}. Contrary to intuition, nearby sources require a higher spin-down power than more distant ones. This is the result of requiring a much softer spectral index for the injected electrons for close-by sources to reproduce the H.E.S.S. spectral index, which implies fewer electrons injected at high energies. At some distance that depends on the magnetic field strength assumed, the distance dominates again over the softening of the spectrum and the $\dot{E}$ increases again.
We find that the pulsar producing the LAES between 1 and 20 TeV should be located between 25--80~pc, with an $\dot{E}$ of $10^{33-34}$ erg/s, with effectively no constraint on the injection spectral index. The different magnetic fields considered do not significantly change the conclusion.

\begin{figure}
\begin{center}
\includegraphics[width=0.49\textwidth]{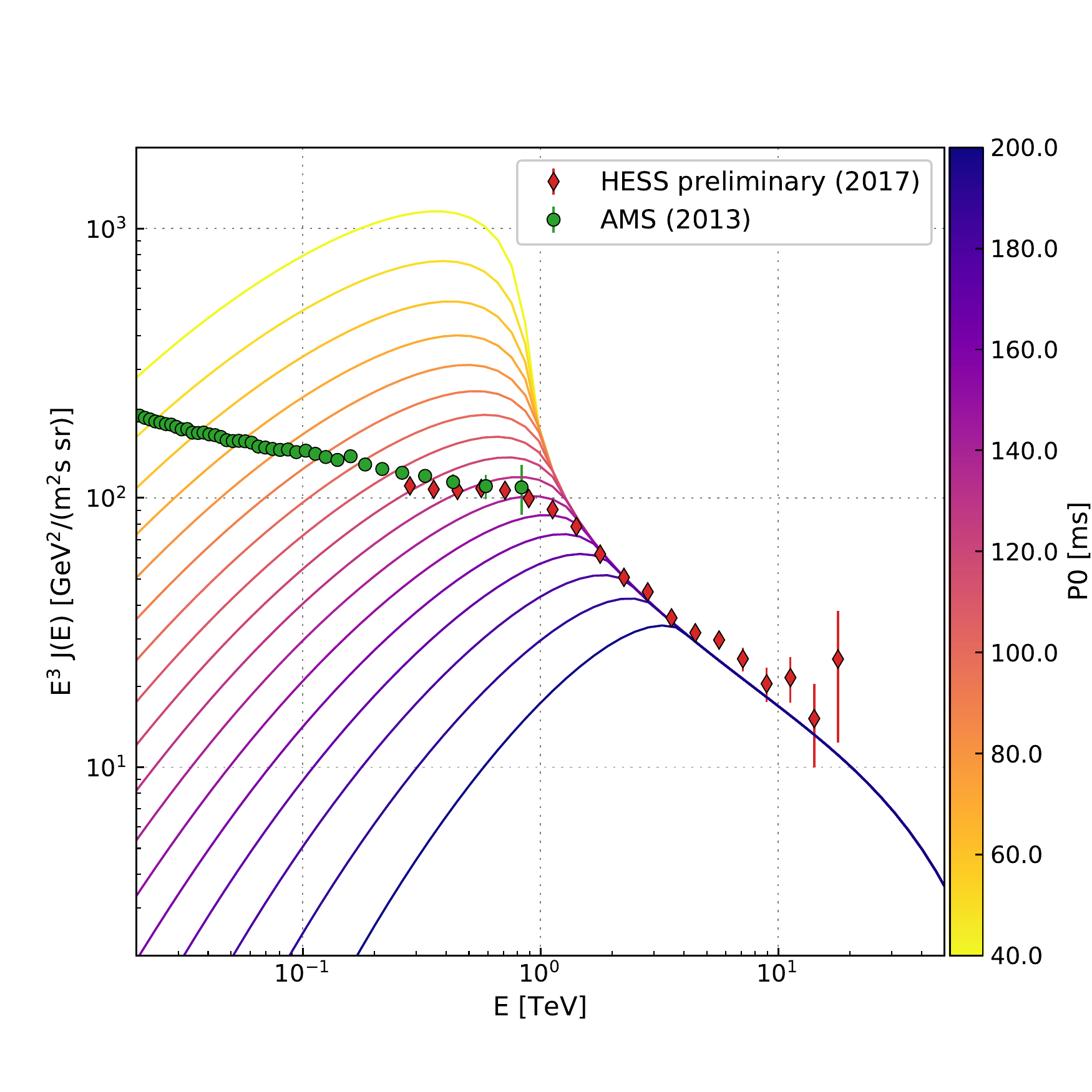}
\caption{All electron spectrum measured at the Earth with AMS~\citep{AMS14} and preliminary H.E.S.S.~\cite{HESS_electrons} data points compared to predictions for a single PWN 
The colored lines correspond to the LAES contribution of a single pulsar 
located at 50~pc, $\dot{E}=1.3\times10^{33}$ erg/s, $\alpha=2.4$, an ISM magnetic field of $B=3\ \mu$G and a range of values for $P_0$.}
\label{fig:all_e}
\end{center}
\end{figure}

In Figure \ref{fig:all_e}, we show the LAES for different $P_0$ values for the example of a pulsar located at 50~pc, with corresponding best fit values of $\alpha$=2.4 and $\dot{E}=1.3\times10^{33}$ erg/s. 
The overall shape of the spectrum at TeV energies is consistent, but it is clear that short initial periods ($P_0 \lesssim 140 $ ms)  overshoot the measured LAES at low energies. However, 
as discussed in \cite{Diffusion_electrons_paper}, those parameters relating to the early evolution of the pulsar ($\tau_0$, $n$ and $P_0$) have a large effect on the prediction in the energy range where the spectrum is uncooled. For the assumed source age and slow diffusion this occurs below $\simeq1$ TeV. 
We note that the assumption of escape without losses becomes increasingly questionable in the early life of a PWN and hence the lowest energy predictions can be seen as upper limits. At higher energies however evolutionary uncertainties are less critical and in the TeV range the single source predictions are more robust. Furthermore, below 1 TeV multiple sources may contribute and a population model is needed to predict the resulting LAES. The requirement for the single source model is therefore consistency with the LAES at TeV energies and to not overshoot the total $e^{+}$ and $e^{-}$ fluxes at lower energies. Birth periods greater than $\sim$160~ms are therefore disfavored as they do not produce sufficient flux around 2~TeV.

\begin{figure}
\begin{center}
\includegraphics[width=0.49\textwidth]{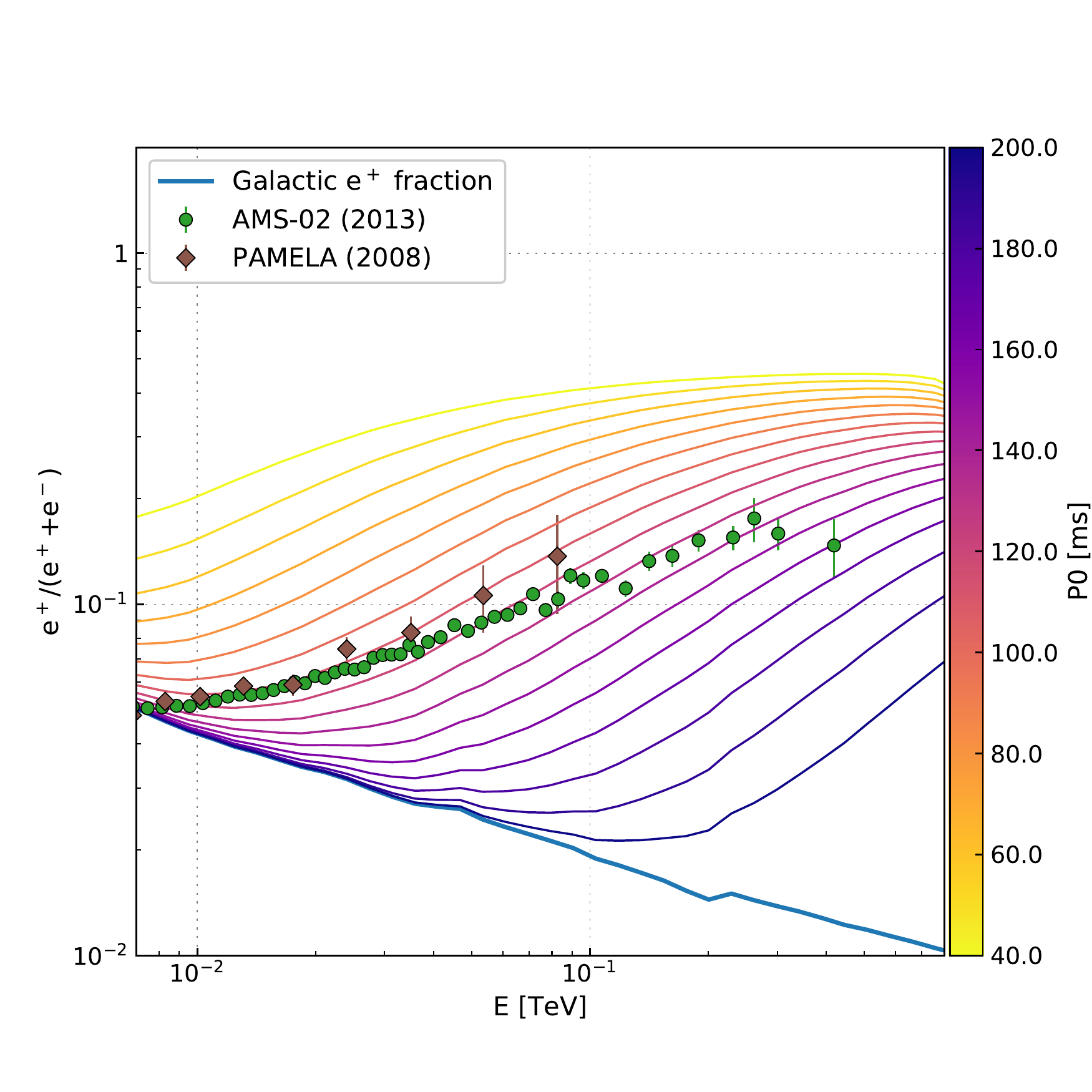}
\caption{Positron fraction measured at the Earth using AMS \citep{AMS_Frac_14} and PAMELA \citep{PAMELA09}. The blue solid line corresponds to the predicted positron fraction due to secondary production from \citep{Moskalenko_Strong_electrons}. The colored lines correspond to the positron fraction with this level of secondaries plus a single PWN as for Figure \ref{fig:all_e}.}
\label{fig:pos_frac}
\end{center}
\end{figure}

Figure~\ref{fig:pos_frac} shows the measured positron fraction and predictions from the single source model. We assume that equal fluxes of electrons and positrons are injected by the hypothetical nearby PWN and a contribution from secondary positrons following~\cite{Moskalenko_Strong_electrons}. As for Figure~\ref{fig:all_e} curves for different values of the pulsar birth period $P_0$ are shown.
Short birth periods ($P_0 \lesssim 140 $ ms) are again formally excluded, but as discussed above this constraint is readily avoided by considering significant losses before escape for those electrons accelerated in the early life of the PWN. As Figure~\ref{fig:npulsars_contributors} illustrates, more than one pulsar is expected to contribute to the LAES and therefore to the positron fraction at energies below hundreds of GeV. However the dominance of a single PWN at high energies necessitates a further rise in the positron fraction beyond the range of current measurements ultimately reaching 0.5, and indeed this is the case for any PWN-dominated model.

\begin{figure}
\begin{center}
\includegraphics[width=0.49\textwidth]{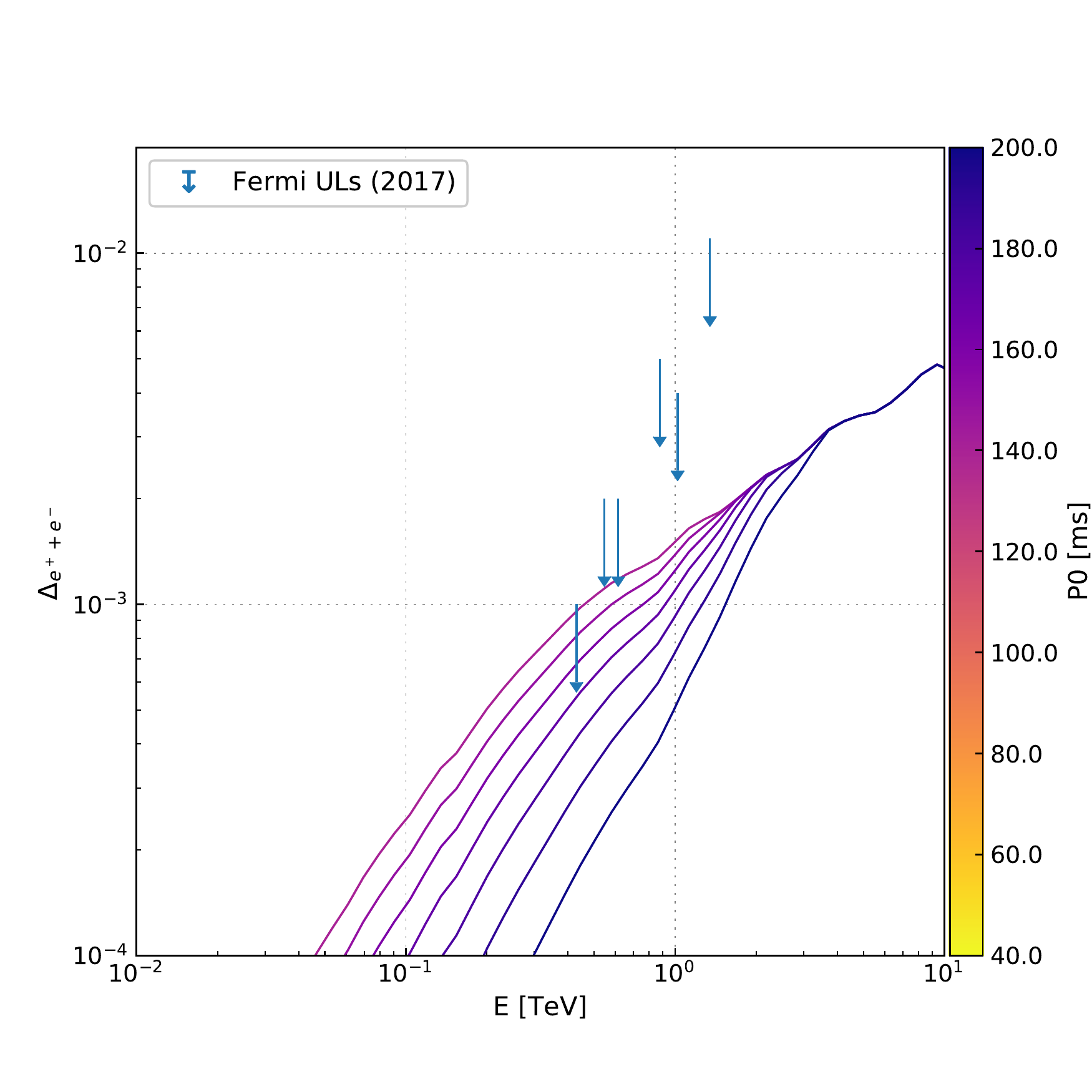}
\caption{Expected anisotropy due to the single local source simulated in Figures \ref{fig:all_e} and \ref{fig:pos_frac}. Fermi upper limits are shown as blue arrows. The color code for the single pulsar contribution is the same as on the aforementioned figures.}
\label{fig:anisotropy}
\end{center}
\end{figure}

The final observational constraint that we consider is the anisotropy of the all-electron flux. Using the LAES, its gradient and the diffusion coefficient, we calculate the anisotropy using equation 3.2 of \cite{Manconi17}. We show the result for curves with a range of $P_0$ that is consistent with the previous constraints. Figure \ref{fig:anisotropy} shows the resulting predictions together with the upper limits established from {\it Fermi-LAT} \cite{Fermi_anisotropy}. 

We note that the conclusions of required birth periods are somewhat modified for values of distance and ISM magnetic field differing from the 50~pc and $3\ \mu$G assumptions made for Figures \ref{fig:all_e}, \ref{fig:pos_frac} and \ref{fig:anisotropy}. However, as Figure~\ref{fig:alpha_vs_distance} shows, solutions exist for a range of distances and magnetic fields provided appropriate choices are made for injected electron spectral index and injection power.


\section{Discussion}

As shown in the previous section the high energy LAES can plausibly be explained by particle acceleration in a single PWN only if it is very local (within $\sim$80~pc) more than $\sim3\times\sim10^5$~years old and with $\dot{E} \sim 10^{33-34}$ erg/s. The measured $\dot{E}$/age relationship~\cite{ATNF} suggests that such $\dot{E}$ values are common for pulsars of around Myr age.
To assess the likelihood of the existence of such a local pulsar we used the source population simulations introduced earlier. These population simulations predict a pulsar within 80 pc in around 5-10\% of the realizations. The existence of such a pulsar implies the occurrence of a recent local supernova, the case for which
is strengthened by studies of $^{60}$Fe transport \cite{Kachelriess15, Schulreich17}. By modeling the formation of the Local Bubble by several supernovae they find that the most recent supernovae must have exploded at a distance of 90-100 pc from the Earth, 2-3 Myr ago with a suitable initial mass to produce a pulsar~\cite{Breitschwerdt16}.

No known pulsar exists with the properties required.
However, the beam fraction of a pulsar is a quite uncertain value and in many cases, for example \cite{Tauris98}, is predicted to be rather small ($<$25\%), in particular for old pulsars. Beaming could potentially lead to a number of pulsars in the local neighborhood which are undetectable by radio observations. These objects, however with their surface temperature of a few million Kelvin may be detectable in soft X-rays by a wide field of view instrument, as for the "Magnificent seven" local neutron stars discovered using ROSAT (for example \cite{ROSAT_NS1, ROSAT_NS2}). Again no candidate sources have been detected within 100~pc, but confirmation and distance measurements of these sources remains difficult due to their often weak optical counterparts. X-ray measurements of a synchrotron pulsar wind nebula as in the case of Geminga \cite{Caraveo03} would also be a way of detecting them, but we note that the X-ray detection of Geminga relied on the early detection of pulsed emission. VHE $\gamma$-ray observations provide an alternative method for detecting PWN and electrons around them (see e.g.~\cite{Klepser_PWN,Geminga_hawc,Linden17}).
Assuming a source powered by a central pulsar with spin-down power 10$^{34}$~erg/s, the VHE $\gamma$-ray flux within 5$^{\circ}$ is more than one order of magnitude lower than that measured by HAWC for Geminga and therefore out of the reach of current or planned VHE facilities.  

\section{Summary}

We have shown that even with the two orders of magnitude lower diffusion coefficient than the conventional expectation inferred from Geminga using HAWC, a single local pulsar wind nebula remains as a feasible source of the highest energy cosmic ray electrons. The required properties of the pulsar powering this nebula are an $\dot{E}\sim 10^{33-34}$ erg/s and an age of $>$300~kyr, at a distance of 
30--80~pc. Population modeling suggests a probability for the existence of such a pulsar is of order 10\% for an age $<$1~Myr. The probability of missing such a pulsar in existing surveys may be rather large due to the beaming fraction of old pulsars and the low surface brightness of the associated VHE $\gamma$-ray nebula. If such a pulsar exists there are good prospects for its discovery in future surveys at (for example) radio or X-ray wavelengths.


\section*{References}
\bibliographystyle{elsarticle-num_etal} 

 \bibliography{./references}


\end{document}